# Demagnetization-Driven Nanoscale Chirality-Selective Thermal Switch

In Hyeok Choi[1,†], Daeheon Kim[1,†], Yeon Jong Jin[1], Seungmo Yang[2,3], Tae-Seong Ju[2], Changsoo Kim[2], Chanyong Hwang[2], Dongbin Shin[1,4,*], and Jong Seok Lee[1,*]

[1]*Department of Physics and Photon Science, Gwangju Institute of Science and Technology (GIST), Gwangju 61005, Republic of Korea*

[2]*Quantum Technology Institute, Korea Research Institute of Standards and Science, Daejeon, 34113, Republic of Korea*

[3]*Department of Physics, Chung-Ang University (CAU), Seoul, 06974, Republic of Korea*

[4]*Max Planck Institute for the Structure and Dynamics of Matter and Center for Free-Electron Laser Science, Luruper Chaussee 149, 22761, Hamburg, Germany*

[†] Equally contributed

[*]Corresponding authors: dshin@gist.ac.kr, jsl@gist.ac.kr

**Abstract**

Chiral-lattice degrees of freedom can offer novel chirality-selective functionalities for thermotronic applications. Chiral phonons, carrying both heat and angular momentum, can emerge through a breaking of chiral degeneracy in the phonon bands, either via an intrinsic chiral crystal structure or by angular momentum transfer from photons or spins. This chiral controllability of the lattice dynamics enables a design of chiral thermo-devices by integrating ferromagnets with chiral materials. Here, we present a nanoscale chirality-selective thermal switch realized using a simple heterostructure composed of ferromagnetic [Co/Pt] multilayers and insulating chiral α-SiO$_2$, where an external magnetic field can control thermal transport properties. Our experimental results based on the magneto-optic thermometry reveal that the thermal conductivity of α-SiO$_2$ exhibits a clear dependence on both the magnetization direction of [Co/Pt] multilayers and the structural chirality of α-SiO$_2$, which is supported well by the first-principles-based molecular dynamic simulations. The magnetization-dependent thermal on/off ratio amounts to 1.07 at room temperature and increases to about 1.2 as temperature decreases to 50 K, due to a reduction of Umklapp phonon-phonon scattering rate in α-$SiO_2$. These findings provide the first experimental demonstration of the nanoscale chirality-selective thermal switch based on the ferromagnetic/chiral material heterostructure, highlighting its potential as a key technology for addressing heat dissipation challenges in nanoscale electronic devices.

Breaking chiral degeneracy in the lattice degrees of freedom gives rise to novel quasiparticles – chiral phonons carrying angular momentum[1-3]. For a long decade, extensive researches have focused on the interaction between chiral degrees of freedom and spins, known as chiral-induced spin selectivity (CISS), facilitating chiral spintronic applications[4-7]. Since CISS relies on interaction between the freely moving electron's spin and the structural chirality, its application is usually restricted to conducting materials. In contrast, chiral phonons can carry angular momentum and energy in both metallic and insulating materials, enabling their applications in insulating materials and hence overcoming the limitations of CISS-based chiral spintronic applications[8,9]. For example, a photo-induced spin current can be induced through the chiral phonon-activated Seebeck effect[9], and chirality-selective heat transfer can be realized via the chiral filtering effect in heterostructures composed of two chiral materials[8], providing additional chiral-phonon-assisted controllability for the spin and thermal transport, respectively.

In chiral materials exhibiting the enantiomeric crystal structure, the chiral degeneracy in the phonon band is inherently lifted[1], resulting in a natural expectation of the chirality-dependent phonon transport in chiral materials. Even in non-chiral materials, chiral phonons have been often excited using circularly polarized light through light-matter interaction[10-12]. However, the optical excitation in the non-chiral materials can induce only transient chiral phonons with short lifetimes, in contrast to the persistent population of chiral phonons in chiral materials. Beyond the direct optical excitation, recent studies have revealed that electron's spin in ferromagnets can efficiently transfer the angular momentum to chiral phonons via spin-phonon coupling[13-15]. Demagnetization in ferromagnets can be induced via optical pumping on an ultrafast timescale, facilitating the transfer of both heat and angular momentum particularly to acoustic chiral phonons, which exhibit much longer lifetimes compared to those for the direct optical excitation[15]. This unique advantage exploiting acoustic chiral phonons opens possibilities for chiral spintronic and thermotronic applications by combining chiral materials and ferromagnets.

Here, we demonstrate chirality-selective heat transfer in heterostructures composed of ferromagnetic (FM) [Co/Pt] multilayers and chiral material (CM) α-SiO$_2$, and a realization of the nanoscale chiral thermal switch. To confirm the chirality-selective heat transfer, we conducted time-resolved magneto-optic Kerr effect (tr-MOKE) measurement, sensitive to the cross-plane thermal transport near the hetero-interface[16]. We measured the apparent thermal conductivities of α-SiO$_2$ ($\kappa_{quartz}$), observing a 7% difference in $\kappa_{quartz}$ by reversing the magnetization *M* in the [Co/Pt] multilayer. Based on our molecular dynamics (MD) simulations, we found that the conservation of the angular momentum in chiral phonons plays an important role in heat propagation from the [Co/Pt] layer into α-SiO$_2$, leading to distinct spatio-temporal temperature profiles in α-SiO$_2$ dependent on the phonon chirality in the [Co/Pt] multilayer. Furthermore, we confirmed that the efficiency of thermal switch can be significantly enhanced at low temperature due to a reduction of phonon-phonon scattering rate. Our novel experimental findings, soundly supported by theoretical calculations, suggest that the chiral degree of freedom in phonons strongly influences thermal transport in chiral materials, offering a new guideline for realizing chirality-selective nano-thermal switches by their simple heterostructuring with ferromagnets.

Chiral α-SiO$_2$, which exhibits an enantiomeric trigonal crystal structure, can have two distinct space groups P3$_1$21 (left-handed chirality) and P3$_2$21 (right-handed chirality)[17] (Section S1, Supporting Information). This chiral crystalline structure leads to the chirality-dependent phonon energy splitting along the [0001] chiral axis. Figure 1 schematically illustrates how the angular momentum transfers from the metallic FM layer to the insulating CM under laser heating. An intense infrared laser beam induces the spin angular momentum transfer to the lattice system generating non-thermalized chiral phonons in the FM layer[13], which become rapidly thermalized via phonon-phonon scattering on a picosecond timescale following the Bose-Einstein distribution[15]. These thermalized chiral phonons can carry both energy and angular momentum to the adjacent CM, where their chirality can be controlled by changing the direction of *M* in FM layer, resulting in chirality-selective heat transfer in the CM.

To confirm the chirality-selective thermal transport in [Co/Pt]/α-SiO$_2$ heterostructures, we obtained κ$_{quartz}$ and thermal boundary conductance ($G_{int}$) with changing the direction of $M_z$ (+$M_z$ or –$M_z$) in the [Co/Pt] multilayer (Fig. 2a), having a single domain state (Section S2, Supporting Information). Here, *z* is along the out-of-plane direction. Figure 2b displays the *M*-direction-dependent thermal responses for the left-handed (left panel) and the right-handed (right panel) α-SiO$_2$, revealing distinct behaviors depending on both the direction of *M* and the chirality of α-SiO$_2$. Here, we define the thermal ratio, reflecting the spin temperature change, as -$V_{in}$/$V_{out}$, where $V_{in}$ and $V_{out}$ are in-phase and out-of-phase signals, respectively, monitored by a lock-in amplifier. In the case of the left-handed α-SiO$_2$, the thermal ratio at +$M_z$ (red line) is slightly larger than that at –$M_z$ (blue line), while this tendency is oppositely changed in the right-handed α-SiO$_2$. It is noteworthy that thermal ratios of non-chiral amorphous SiO$_2$ exhibit no significant difference depending on the direction of $M_z$ (Sections S3 and S4, Supporting Information). These *M*-direction- and chirality-dependences can be seen more clearly from the difference in thermal ratios of opposite directions of $M_z$ (ΔRatio), as displayed in the top panel of Fig. 2c. Therefore, we can rule out artificial errors in the thermal ratio from reflectance changes. The bottom panel shows corresponding sensitivity functions for several thermal parameters defined as ln*R*/ln*α*, where *R* and *α* are thermal responses and thermal parameters, respectively[18]. The sensitivity functions provide qualitative information on how thermal responses are changed with respect to a modulation of individual thermal parameter, thereby identifying which thermal parameters would contribute to the thermal ratio and ΔRatio (Sections S5, Supporting Information). Here, we consider four parameters, the thickness ($d_{Co/Pt}$, blue) of the [Co/Pt] multilayer, its thermal conductivity ($\kappa_{Co/Pt}$, black), $G_{int}$ (green), and $\kappa_{quartz}$ (blue). Among them, whereas $d_{Co/Pt}$, $G_{int}$, and $\kappa_{quartz}$ show meaningful sensitivity, the sensitivities of $d_{Co/Pt}$ and $\kappa_{quartz}$ exhibit an exponential decay, resembling the time-dependent behavior of ΔRatio. This suggests that $d_{Co/Pt}$ and $\kappa_{quartz}$ primarily contribute to changes in the thermal ratio. Since $d_{Co/Pt}$ is a fixed parameter from the beginning, we conclude that $M$- and chirality-dependent thermal ratios should be contributed to by corresponding changes of $\kappa_{quartz}$, demonstrating the existence of chirality-selective thermal transport [Co/Pt]/α-SiO$_2$ heterostructures which can be controlled by the direction of $M_z$.

By fitting the thermal ratio using the thermal diffusion equation[18], we obtained the $M$-direction-dependent $\kappa_{quartz}$ for the left-handed, the right-handed, and the non-chiral SiO$_2$. As shown in Figs. 2d-f, $\kappa_{quartz}$ values obtained at several different positions clearly demonstrate the chirality-selective switchable thermal functionality in the FM/CM heterostructure. $\kappa_{quartz}$ of the chiral α-SiO$_2$ and the non-chiral SiO$_2$ are approximately 11 Wm$^{-1}$K$^{-1}$, and 1.5 Wm$^{-1}$K$^{-1}$, consistent with the reported values for crystalline and amorphous SiO$_2$, respectively[19]. Importantly, $\kappa_{quartz}$ of the chiral α-SiO$_2$ exhibits statistically distinct values by about 0.7±0.1 Wm$^{-1}$K$^{-1}$ depending on $M_z$, and such $M$-direction-dependences are symmetrically reversed for the chiral α-SiO$_2$ having the opposite handedness. It should be noted that the non-chiral SiO$_2$ exhibits negligible $M$-direction-dependence of $\kappa_{quartz}$ (Fig. 2f). Furthermore, we confirmed that $G_{int}$ also exhibits both $M$- and chirality-dependence (Sections S6, Supporting Information).

We can define a thermal on/off ratio as $\kappa_{on}/\kappa_{off}$, where $\kappa_{on}$ and $\kappa_{off}$ are larger and smaller values of $\kappa_{quartz}$ depending on the $M_z$ direction for the given chirality of α-SiO$_2$, and it is estimated as about 1.07 at room temperature. The chirality of demagnetization-induced chiral phonons in the [Co/Pt] multilayers follows the direction of $M_z$, exhibiting positive or negative angular momentum. Therefore, the $M$-direction-dependent $\kappa_{quartz}$ of the chiral α-SiO$_2$ suggests that the efficiency of heat transfer across the heterostructures is much higher when the angular momentum of chiral phonons and the chirality of α-SiO$_2$ match each other. To further confirm the role of ultrafast demagnetization in the thermal transport, we obtained *M*-direction-dependent thermal on/off ratio for the same heterostructure with an additional 70 nm Al transducer on it, where the pumping beam is fully absorbed. In this case, the thermal gradient may induce the weaker demagnetization in the FM layer compared to the direct excitation. Indeed, the thermal on/off ratio is estimated as about 1.02 for the FM/CM heterostructure with Al transducer (Section S7, Supporting Information), which is significantly smaller than in the direct excitation case. Consequently, the apparent κ$_{quartz}$ with respect to $M_z$ clearly demonstrates the chiral phonon diode effect in α-SiO$_2$ and realizes the chirality-selective thermal switch with the FM/CM heterostructure which can be controlled by the external magnetic field. We note that the thermal on/off ratio of our FM/CM heterostructure is relatively small compared to previously reported nanoscale thermal switches[20]. For instance, [Co/Cu] multilayers exhibit *M*-dependent giant magnetothermal resistance (~100 %), attributed to electron's spin heat accumulation at the interface[21]. On the other hand, our FM/CM heterostructures can exhibit the thermal switch functionality exploiting the thermal diode effect in the insulating CM where the effective magnetic moment of chiral phonons depends on their propagation direction[8]. This highlights the potential application of FM/CM heterostructures, for example, for the thermal regulation with an electric insulation efficiently secured.

First-principles-based classical MD simulations further provide a comprehensive understanding of chirality-selective thermal transport in FM/CM heterostructures. To maintain consistency with the experimental configuration, we employed a bi-layer structure of 4 nm Co and 165 nm right-handed α-SiO$_2$, as shown in Fig. 3a. Chiral phonons in the Co layers were induced by enforcing clockwise (CL) or counter-clockwise (CC) rotations of Co atoms with 1 THz frequency and 2 K kinetic energy. The α-SiO$_2$ layers were prepared with 1 K kinetic energy with NVT ensemble at time *t* = 0. Figures 3b and 3c exhibit spatiotemporal heat propagation from the Co layers to the $SiO_2$ layers with the CL and CC rotations of Co atoms, respectively. In addition, $T$ difference ($\Delta T$) obtained between the CL and CC rotations is depicted in Fig. 3d, denoted by CL–CC. At the early stage ($t < 30$ ps), the CL rotation of Co atoms gives a slower heat propagation toward right-handed α-$SiO_2$ than the CC rotation, resulting in negative $\Delta T$ (red area in Fig. 3d). At the saturated region ($t > 40$ ps), however, $T$ with the CL rotation shows the more efficient heat transfer than the CC rotation. Accordingly, the sign of $\Delta T$ becomes positive over time (blue area in Fig. 3d). This chirality-dependent heat propagation in the α-$SiO_2$ layer implies the chiral phonon diode effect, hence confirming a possible realization of the thermal switch effect in the FM/CM heterostructures.

The angular-momentum-polarized phonon structure of α-$SiO_2$ plays a critical role in the chiral diode effect in α-$SiO_2$[8]. To understand the chirality-selective thermal switch based on this effect, we evaluated the $z$-component of phonon angular momentum $L_z$ for each $q$-vector from the calculated phonon dispersion ($\Gamma$ to $-A$) of right-handed α-$SiO_2$ (see Fig. 3e). We note that phonons in $\Gamma$ to $-A$ mainly contribute to the thermal transport along the negative $z$ direction in our coordinate. Below 4.5 THz, there are three acoustic branches, which have positive (red), zero (grey), and negative (blue) $L_z$, suggesting that thermal transport in the α-$SiO_2$ layer should be influenced by the chirality of phonons induced in the Co layers. Considering the heat transfer from the coherent 1 THz phonon, which is the rotation frequency of Co atoms, the interlayer phonon-to-phonon transfer can occur resonantly to the α-$SiO_2$ layer (Section S8, Supporting Information). On the other hand, the α-$SiO_2$ layer has both positive and negative $L_z$ phonons at 1 THz along $\Gamma$ to $-A$. At the early stage of dynamics, for this reason, the higher group velocity of CC phonon introduces the faster heat propagation, as presented in Figs. 3b and 3c. The dynamics at later times ($t > 50$ ps), however, are determined by an additional contribution from the density of states of CL and CC phonons. To quantify this behavior, we calculated the integrated angular momentum $L_z^{\text{int}}(\omega) = \int_{-A}^{\Gamma} g^\lambda(\omega, q) L_z(q) dq$, where $g^\lambda(\omega, q)$ is a partial density of states for $\lambda$-branch at $q$-vector. As shown in Fig. 3f, $L_z^{\text{int}}$ is negative at 1 THz, and this indicates the more efficient heat propagation from the Co layers to the α-SiO$_2$ layers when the Co atoms exhibit the CL rotation, corresponding to negative $L_z$. This description is in good agreement with our MD simulation result shown in Fig. 3d. However, if we induce chiral phonons in the Co layers by rotating Co atoms with 3.5 THz, $L_z^{\text{int}}$ now have a positive value at this frequency, resulting in an efficient heat transfer into the α-SiO$_2$ layer with the CC rotation having a positive $L_z$ (Section S9, Supporting Information). In our experimental results, discussed in Fig. 2, κ$_{\text{quartz}}$ exhibits a larger value at $-M$ in the right-handed α-SiO$_2$, where the chiral phonons can be generated with a negative $L_z$. This indicates that the broken chiral degeneracy in the phonon bands, particularly in the low-frequency region ($< 2$ THz), can contribute to chirality-selective thermal transport in [Co/Pt]/α-SiO$_2$ heterostructures.

The crucial role of chiral phonons in α-SiO$_2$ in determining the chiral phonon diode effect suggests that the performance of a chiral thermal switch can be further tuned by a temperature ($T$) variation, accompanied by changes in phonon dynamics. Figure 4a displays $T$-dependent κ$_{\text{quartz}}$ of the left-handed (left panel) and the right-handed (right panel) α-SiO$_2$ obtained at $+M_z$ (red symbol) and $-M_z$ (blue symbol). κ$_{\text{quartz}}$ exponentially increases as $T$ decreases and shows a clear separation depending on the direction of $M$, which exhibits an opposite tendency between the left-handed and the right-handed SiO$_2$ (Section S10, Supporting Information). This suggests that the chiral thermal switch operates even down to 50 K. As presented in Fig. 4b, thermal on/off ratios obtained with 10 nm (square) and 20 nm (circle) [Co/Pt] multilayers show strong $T$ dependences, exhibiting a fourfold enhancement as $T$ decreases from 300 K to 50 K. We note that the performance of the thermal switch less depends on the thickness of [Co/Pt] multilayers. We compared $T$-dependent κ$_{\text{quartz}}$ with the Callaway model prediction[22] considering the Umklapp process as the dominant phonon scattering process in crystalline bulk α-SiO$_2$[23], and found that they are in good agreements with each other. As displayed with dashed lines in Fig. 4, κ$_{quartz}$ shows the maximum value, corresponding to the Umklapp peak, around 20 K, consistent with the previous report[23]. At $T > 20$ K, namely above the Umklapp peak, the phonon scattering rate is approximately given by $\sim C e^{-\alpha \theta_\mathrm{D}/T}$, where $C$, α, and $\theta_\mathrm{D}$ are the strength of the Umklapp process, a phonon dispersion-related constant, and the Debye temperature, respectively. This implies that the phonon scattering rate is exponentially reduced as $T$ decreases above the Umklapp peak, and the phonon angular momentum conductivity in α-SiO$_2$ would be enhanced accordingly[23]. This can affect the efficiency of the chiral phonon diode effect, and finally contribute to the increase of the thermal on/off ratio upon cooling. We note that the thermal on/off ratio can be further modulated by changing the thermal penetration depth, highlighting the crucial role of phonon properties in SiO$_2$ in the chirality-dependent thermal transport (Section S11, Supporting Information).

In conclusion, we experimentally observed the ultrafast demagnetization-driven chiral phonon diode effect in α-SiO$_2$, and demonstrated the chirality-selective, switchable thermal functionality operating at room temperature by its heterostructuring with ferromagnetic [Co/Pt] multilayers. Both optical excitation and thermal gradients induce the demagnetization in the [Co/Pt] multilayers, generating chiral phonons whose angular momentum sign is determined by the initial magnetization direction. These chiral phonons transfer to the adjacent chiral α-SiO$_2$, leading to chirality-selective thermal conductivity via the chiral phonon diode effect. We found that the thermal on/off ratio is 1.07 for the direct optical excitation at room temperature, and it can be enhanced to 1.2 at 50 K due to the reduction of phonon-phonon scattering rate in chiral α-SiO$_2$. Our extensive first-principles-based calculations provide a comprehensive understanding of the angular momentum transfer process in FM/CM heterostructures based on the phonon dispersion with phonon angular momentum encoded. Therefore, we believe our findings can provide a blueprint for a prototype nanoscale thermal switch and thermal diode that can be implanted using the simple heterostructure, with broad applicability to nanoscale electronic devices for addressing heat dissipation challenges.

# Method

## Sample preparation

The ferromagnetic layer was prepared as multilayered structures, Ta (2 nm)/Pt (5 nm)/[Co (0.7 nm)/Pt (1 nm)]$_6$/Pt (2 nm) for the 20 nm sample and Ta (2 nm)/Pt (3 nm)/[Co (0.5 nm)/Pt (0.5 nm)]$_4$/Pt (1 nm) for the 10 nm sample. The subscript indicates a repetition of the corresponding Co/Pt bilayers. The substrates used in the experiments include Si/amorphous SiO$_2$, and (0001)-oriented chiral α-SiO$_2$ (*MTI corp.*). These multilayer stacks were deposited using a magnetron DC sputtering system, with a base pressure of $8 \times 10^{-9}$ Torr.

## Time-resolved Magneto-optic Kerr effect measurement

To determine thermal parameters for [Co/Pt] multilayers/α-SiO$_2$ heterostructures, we employed the time-resolved magneto-optic Kerr effect (tr-MOKE) technique. A pulsed laser beam (VISION-S, *Coherent*) with a repetition rate of 80 MHz was separated into pump and probe beams using a polarizing beam splitter. Both the pump and probe beam were focused on the sample surface by a 5× objective lens (*Thorlabs*) at a normal incidence, and their beam sizes were approximately 12 μm (1-σ of Gaussian function). The pump with a power 20 mW increases the sample temperature by about 1 K, and the probe beam with a power of 2 mW probes the pump-induced change in MOKE signals modulated with dual-frequencies 50 kHz using a photoelastic modulator (*Hinds*) and 10 MHz using an electro-optical modulator (*Conoptics*). The pump beam was optically blocked in front of a high-bandwidth photo-diode detector (*Thorlabs*) using optical shortpass and longpass filters with the two-tint method. The transient MOKE signals were fitted by a thermal diffusion equation considering the thermal resistor model to obtain the thermal conductivity of α-SiO$_2$ and the thermal boundary conductance between the [Co/Pt] transducer and the α-SiO$_2$ substrate. We note that the thermal conductivity and the heat capacity of [Co/Pt] transducer and α-SiO$_2$ were adopted by previous reports[24].

**Molecular dynamics simulation with machine learning force field from the first-principle calculations**

To investigate the electronic structures and phonon structures, we performed density functional theory (DFT) calculations using the Vienna *ab-initio* simulation package (VASP)[25-27]. Perdew-Burke-Ernzerhof[28] type generalized gradient approximation was employed to describe the electron exchange-correlation potential. The projector-augmented wave method was used to describe the core-level state under the pseudopotential scheme[29]. The plane wave basis set is used with 400 eV energy cutoff. Monkhorst-Pack[30] *k*-point sampling is employed, 9×9×7 *k*-points mesh is used for SiO$_2$, and 9×9×1 *k*-points mesh is used for 3 layers of SiO$_2$ and 2 layers of Co with a vacuum slab up to 40 Å. To obtain a machine learning force field for simulating classical molecular dynamics, we performed DFT molecular dynamics of a merged SiO$_2$ and Co system with enough vacuum space. To accumulate the training data points, the NVT ensemble was computed at temperatures of 50, 300, and 500 K using the Verlet algorithm. The Born-Oppenheimer molecular dynamics was performed up to 1 ns with a 1 ps time step for each condition. We employed SIMPLE-NN package for machine learning force field (MLFF) training[31]. We used a 90-60-60-1 network which was optimized by Adam optimizer with a 0.001 learning rate and an 8 batch size for 1200 epochs. We employed classical molecular dynamics using the LAMMPS[32] to compute the heat transfer from Co to SiO$_2$ in the larger and longer scales. The simulation system consists of Co 1×1×4 supercells on the SiO$_2$ 1×1×165 supercells with 25 Å vacuum space along the *z* axis. Co was forced to have a circular motion in the clockwise or counterclockwise direction with 2 K kinetic energy. The NVE ensemble was used for the dynamics of SiO$_2$ with the circular motion of Co. The initial temperature of SiO$_2$ was set to 1 K under the NVT annealing.

## Acknowledgements

J.S.L., I.H.C., and Y.J.J. acknowledge the support from the Korea government (MSIT) No. RS-2022-NR070254. I.H.C. was supported by the National Research Foundation of Korea (NRF) grant funded by the Korea government (MSIT) (No. RS-2024-00351794). D.S. and D.K. was supported by the National Research Foundation of Korea (NRF) grant funded by the Korea government (MSIT) (No. RS-2024-00333664) and the Gwangju Institute of Science and Technology (GIST) research fund (Future-leading Specialized Research Project, 2025). The computational work was supported by the National Supercomputing Center with supercomputing resources including technical support (KSC-2024-CRE-0124). This research was supported by the National Research Council of Science & Technology (NST) grant by the Korean government (MSIT) (No. GTL24041-000), by the National Research Foundation of Korea (Grant No. RS-2022-NR068190). This research was supported by K-CHIPS (Korea Collaborative & High-tech Initiative for Prospective Semiconductor Research) (RS-2025-02310372, 25068-15FC) funded by the Ministry of Trade, Industry & Energy (MOTIE, Korea).

## Contributions

I.H.C. and D.K. contributed equally to this work. J.S.L. and D.S. supervised the project. I.H.C., Y.J.J., and J.S.L. performed TDTR measurements and analyzed the data. I.H.C. and J.S.L. performed SHG, and MOKE measurements and analyzed the data. S.Y., T.S.J., and C.H. prepared thin films. C.K. performed Raman measurements. D.K. and D.S. performed numerical calculations. I.H.C., D.K., D.S., and J.S.L. wrote the manuscript. All the authors discussed the results, implications, and commented on the manuscript.

# Figures

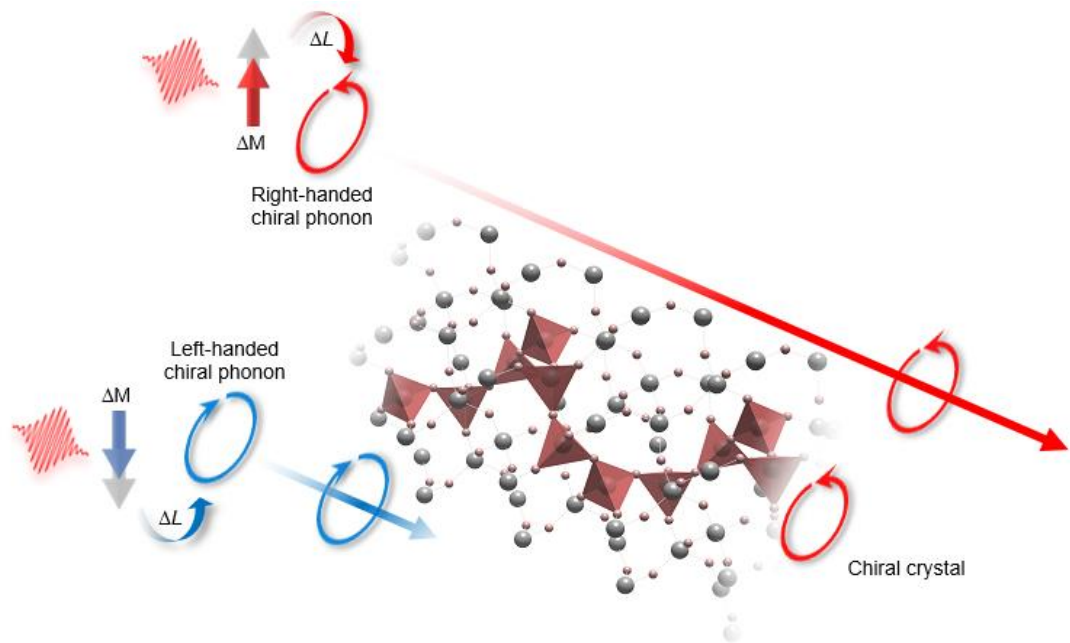

**Figure 1. Chirality-selective thermal switch realized using ferromagnet/chiral material heterostructures.** Optical excitation can lead to the ultrafast demagnetization ($\Delta M$) in the ferromagnet with a loss of the spin angular momentum ($-\Delta L$). The angular momentum transfers to the lattice system generating chiral phonons, which propagate into the adjacent chiral material. When the chirality of the chiral phonons matches that of the chiral material, the chiral phonons can travel more efficiently in the chiral material. This results in the magnetization-dependent thermal conductivity in the ferromagnet/chiral material heterostructures realizing the chirality-selective thermal switch functionality.

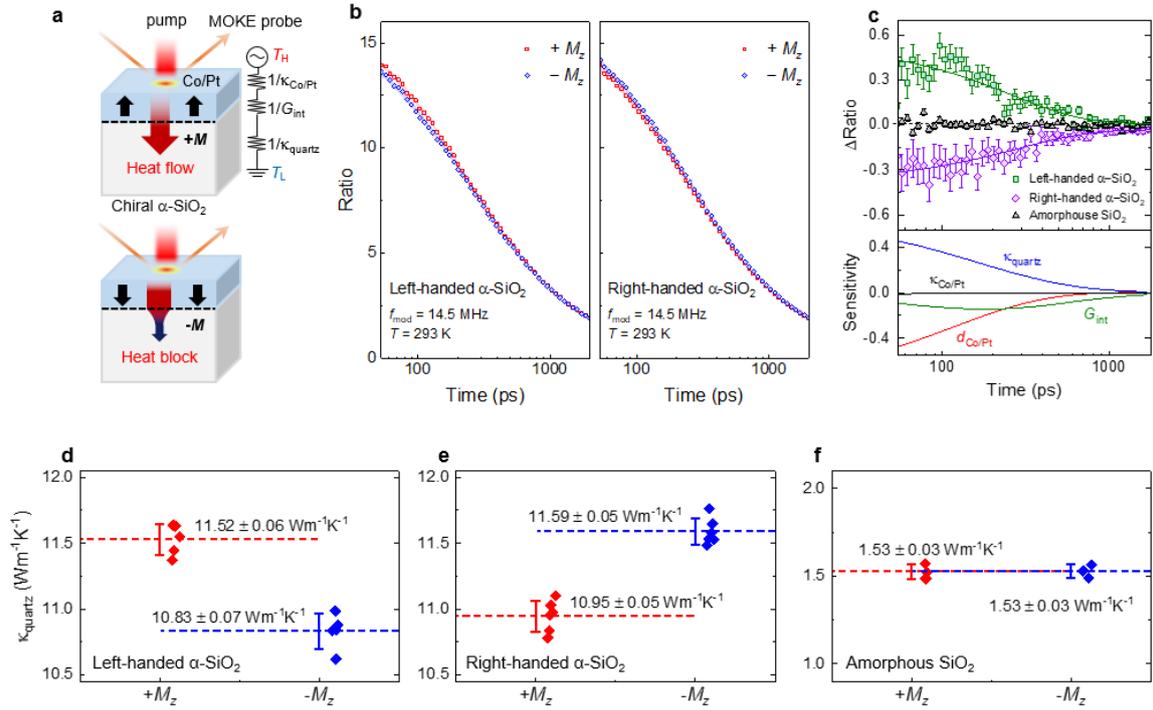

**Figure 2. Magnetization-dependent thermal conductivity of α-SiO$_2$ in [Co/Pt]/α-SiO$_2$ heterostructures.** (a) Magneto-optic thermometry measurements by tracing spin temperature with the magneto-optic Kerr effect (MOKE) technique. We obtained the thermal conductivity with changing the direction of magnetization along the up (top) and down (bottom) directions. (b) Magnetization-direction-dependent thermal ratio obtained with the pump modulation frequency of 14.5 MHz at room temperature for left-handed and right-handed α-SiO$_2$. (c) (top) Difference in thermal ratios obtained at the opposite direction of magnetization for left-handed (green), right-handed (purple), and non-chiral amorphous (black) SiO$_2$. (bottom) Sensitivity functions for several thermal parameters as a function of time. (d-f) Magnetization-dependent thermal conductivity for left-handed (left), right-handed (middle), and non-chiral amorphous (right) SiO$_2$. κ$_{quartz}$ taken at several random positions are displayed in each configuration of magnetization and handedness.

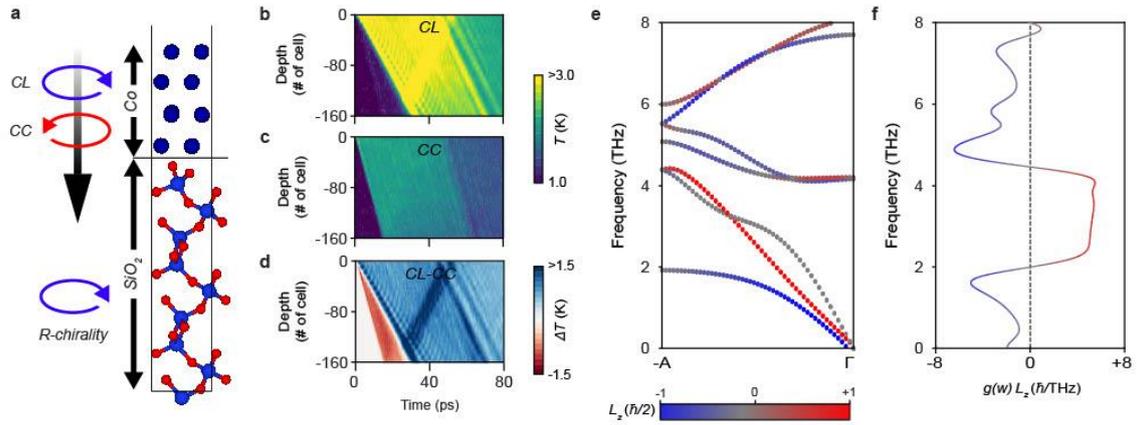

**Figure 3. Molecular dynamics (MD) simulation for chirality-dependent thermal transport in Co/α-SiO$_2$ heterostructures.** (**a**) Illustration of the simulation configuration. The heterostructure was composed of 5.5 nm Co (top layer) and 85 nm right-handed α-SiO$_2$ (bottom layer), relaxed for all regions at 1 K with a total cell length of 91.6 nm. (**b, c**) Spatiotemporal heat propagation in the right-handed α-SiO$_2$ as functions of depth ($d$) and time ($t$) induced by coherent rotation of Co atoms with 1.0 THz in the clockwise (CL) and counterclockwise (CC) direction, respectively. (**d**) Temperature distribution difference between CL and CC motion-driven dynamics ($\Delta T(t, d) = T^{CL}(t, d) - T^{CC}(t, d)$). (**e**) Phonon dispersion along –A to Γ line with the angular momentum of phonon indicated at each $q$-vector. (**f**) Integrated angular momentum of phonon along –A to Γ line with respect to phonon frequency.

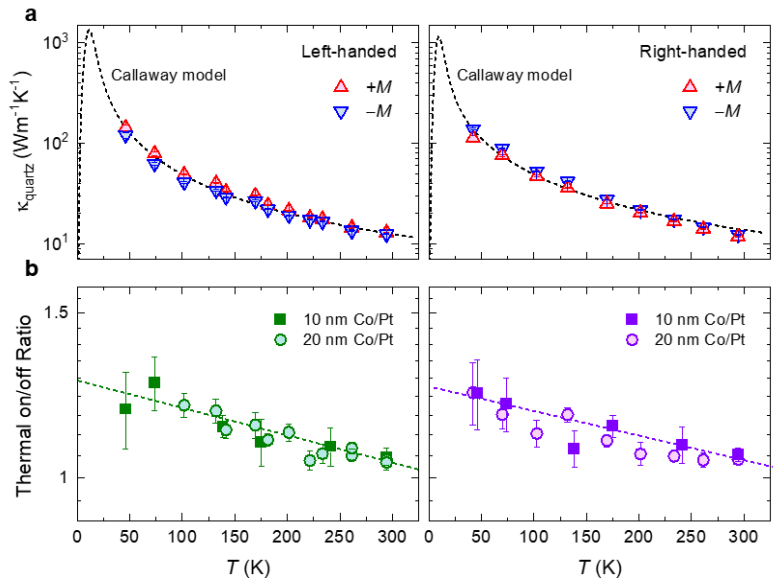

**Figure 4. Temperature-dependent chiral-selective thermal transport in [Co/Pt]/α-SiO$_2$ heterostructures.** **(a)** Temperature- and magnetization-dependent thermal conductivity for left-handed (left) and right-handed (right) α-SiO$_2$. Dashed lines indicate the thermal conductivity calculated from the Callaway model considering the Umklapp scattering process. **(b)** Temperature-dependent thermal on/off ratios for left-handed (left) and right-handed (right) α-SiO$_2$ with 10 nm (square) and 20 nm (circle) Co/Pt multilayers.